%% file: manuscript.tex
\documentclass[12pt]{iopart}
\usepackage{iopams}  
\usepackage{epsfig}
\usepackage{float}
\usepackage{color}
\usepackage{amssymb}
\usepackage{bm}
\usepackage[utf8]{inputenc}
\usepackage[colorlinks=true,linktoc=page,linkcolor=magenta,citecolor=magenta]{hyperref}

\input manuscript_macro

\begin{document}

\title[Persistent Current of Dirac fermions with impurities]{Persistent current of relativistic electrons on a Dirac ring in presence of impurities}

\author{Sumit Ghosh}

\address{S N Bose National Centre for Basic Sciences, \\
Salt Lake, Kolkata 700098, India \footnote{Present address: PSE Division, KAUST
Thuwal 23955-6900, KSA}}
\ead{sumit.ghosh@kaust.edu.sa}

\author{Arijit Saha}

\address{Department of Physics, University of Basel\\
Klingelbergstrasse 82, CH-4056 Basel, Switzerland}
\ead{arijit.saha@unibas.ch}

\begin{abstract}
We study the behavior of persistent current of relativistic electrons on a one dimensional ring in presence of attractive/repulsive scattering potentials. 
In particular, we investigate the persistent current in accordance with the strength as well as the number of the scattering potential. We find that in presence 
of single scatterer the persistent current becomes smaller in magnitude than the scattering free scenario. This behaviour is similar to the non-relativistic case. 
Even for a very strong scattering potential, finite amount of persistent current remains for a relativistic ring. In presence of multiple scatterer we observe that 
the persistent current is maximum when the scatterers are placed uniformly compared to the current averaged over random configurations. However if we increase the 
number of scatterers, we find that the random averaged current increases with the number of scatterers. The latter behaviour is in contrast to the non-relativistic case.
\end{abstract}

\maketitle

%--------------------------------------------------
\section{Introduction \label{sec:intro}}
%---------------------------------------------------
Persistent current (\pc) in a normal metallic ring has drawn significant attention in the last few decades~\cite{Butt1,Land,Gefen1,Gefen2,Alex,Weisz,Rab,Butt2,Janie,Kulik}. 
Due to the recent advancement in fabrication techniques, it has now become possible to observe this current experimentally~\cite{Levy,Maily,Klee,Jayich} in normal metal rings. 
Such current is also observed in carbon nanotube rings~\cite{Szopa,Latil,Chena} in which the low energy spectrum of electrons require a relativistic description. 
There are a large number of intriguing literatures on the energy levels and \pcd of 
relativistic fermions~\cite{Papp,Su,Cayssol} involving the condensed matter systems like ~graphene~\cite{Ino,Castro,Recher,Zar,Peet,Huang} 
and topological insulator~\cite{Mich} in which the low energy spectrum is also described by Dirac Hamiltonian. 
For large particle mass the prediction of both relativistic and nonrelativistic theory is similar, but for small particle mass they differ significantly. 
Some remarkable difference can be observed in case of scattering and barrier penetration at this mass regime. Although the nature of \pcd is well studied for 
the condensed matter systems, there are still several aspects of relativistic fermions which needs a further detailed analysis regarding scattering by a 
potential barrier/well on a ring geometry. 

Motivated by the above mentioned facts, in this article we investigate the nature of energy and \pcd spectra of a relativistic fermion on a \od ring in presence of 
both attractive and repulsive scattering potentials. Here we consider finite width as well as different strengths and configurations for the scatterers along the ring. 
We adopt the relativistic Kronig-Penney model~\cite{McKeller} to carry out our analysis which has been widely used in recent years for studying the effect of 
potential barriers in graphene~\cite{Barb,Barb2,Barb3,Masir,Titov,Matu}. 
%We work here with a $2\times2$ Dirac equation which is quite sufficient in describing a graphene monolayer for a single valley. 
Also the model we adopted here is sufficient to describe the low energy spectrum of a monolayer graphene with single valley and finite mass gap.
Previously such model has been considered in Ref.~\cite{Peet} where they have considered quantum rings made out of single layer 
graphene described by the Dirac Hamiltonian with a finite mass term. Motivated by their model, in our work we consider Dirac Hamiltonian with a finite mass gap 
(similar to single valley of a monolayer graphene with a constant mass gap) and the effects of single and random impurities on energy spectrum and persistent current therein. 
We also assume that, in our model, impurities do not break the valley degeneracy such that considering single valley is a legitimate approximation.

Although the model is developed for relativistic particles, but for high particle mass it can predict the nonrelativistic nature as well. 
Our main focus is to investigate the behavior of \pcd  with different configurations and strengths of the scatterers. We also investigate the variation of the maximum \pcd with single 
scattering potential and show how the rate of change of \pcd varies with respect to the strength of scattering potential for different mass gap. 
Finally we make a comparison among random configurations of multiple scattering potentials along the ring and show that if the fraction of ring covered by 
the impurities remain the same, \pcd after the disorder average becomes larger in magnitude with the increase of number of impurities. 
The latter behaviour is in contrast to the non-relativistic case.

The organization of the rest of the article is as follows. In Sec.~\ref{sec:model} we present a generic description of our model for the quantum ring with 
relativistic electrons in presence of scattering potentials. 
In Sec.~\ref{sec:rel} we analyse the spectrum and \pcd for a relativistic quantum ring in presence of single scatterer of different strenghts. 
In Sec.~\ref{sec:scattering} we discuss the \pcd in presence of uniform as well as random configurations of scatterers after disorder average.
Finally in Sec.~\ref{sec:conclusion}, we present our summary and conclusion.

%--------------------------------------------------
\section{Model and method \label{sec:model}}
%---------------------------------------------------
A quantum ring presents an ideal realization of an infinite lattice with periodicity $\rm 2\pi R$, $\rm R$ being its radius. Due to this periodicity one can use the Kronig-Penney 
(\kp) model~\cite{Kittel} to study the effect of impurities in such a ring. In our analysis we incorporate the relativistic version of the \kpd model. This model was first 
introduced to study the behaviour of quarks in a periodic nuclear lattice~\cite{McKeller}. Here we have generalized this approach for multiple scatterer in an unit cell. 
The method basically involves two steps$-${\textsl{(i)}}~finding the solution of Dirac equation in different regions of the ring 
and {\textsl{(ii)}} connecting the solutions using the transfer matrices. Here we model our impurities within the relativistic ring as rectangular well/barrier 
with finite width and strength. For vanishing scattering potential strength, our results match exactly with the analytical predictions~\cite{Peet}.

%--------------------------------------------------------------------------------------------------
%
%--------------------------------------------------------------------------------------------------
\begin{figure}[ht]
\centering
\epsfig{file=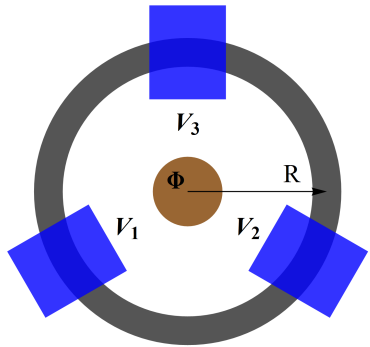,scale=0.4}
\caption{(Color online)~Schematic of our geometry where a magnetic flux $\Phi$ is threaded through a ring of radius $R$ with relativistic electrons. Here $V_{1}$, $V_{2}$ and $V_{3}$ 
etc correspond to the scalar impurities modelled as potential wells/barriers placed along the ring.}
\label{geometry}
\end{figure}
%-----------------------------------------------------------------------------------------------------
%
%-----------------------------------------------------------------------------------------------------
Here we briefly discuss our model which starts with the Dirac equation for electrons on a \od quantum ring. The model Hamiltonian is sufficient to describe the low energy spectrum of
a graphene monolayer for single valley with a constant mass gap~\cite{Peet}. The Dirac equation is given by

\begin{eqnarray}
\left( \begin{array}{cc} \delta & -i\frac{\hbar c}{R} e^{-i\varphi}\left(-\frac{1}{2}-i \frac{\partial}{\partial \varphi}\right) \\ -i\frac{\hbar c}{R} e^{i\varphi} \left( -\frac{1}{2} + i \frac{\partial}{\partial \varphi} \right) & -\delta \end{array}\right)
\left( \begin{array}{cc} \zeta_1 \\ \zeta_2 \end{array}\right) = E \left( \begin{array}{cc} \zeta_1 \\ \zeta_2 \end{array}\right) \ ,
\end{eqnarray}

where $R$ is the radius of the ring and $\zeta_1$ and $\zeta_2$ are scalar functions of the angular variable $\varphi$ satisfying the normalization condition 
$|\zeta_1|^2+|\zeta_2|^2=1$. Here $\delta$ corresponds the mass term or the band gap in case of condensed matter system. The Dirac equation within the square well/barrier 
can be obtained by replacing the energy $E$ with $E-V_0$, where $V_0$ is the strength of the well/barrier. 
Hence we obtain two sets of linearly independent solutions characterized by a spatial part $e^{il_+\varphi}$ and $e^{il_-\varphi}$ which can be written as

\begin{eqnarray}
&&\psi^\pm_l(\varphi) = \left( \begin{array}{c} \zeta_1^\pm \\ \zeta_2^\pm \end{array} \right) 
= \left( \begin{array}{c} 1 \\ i \kappa_\pm e^{i \varphi} \end{array} \right) e^{il_\pm \varphi}\ ,
\end{eqnarray}
where $\kappa_\pm = \frac{\hbar c}{R} \frac{1/2 + l_\pm}{E+ \delta}$. 

%--------------------------------------------------------------------------------
%
%--------------------------------------------------------------------------------
\begin{figure}[h]
\centering
\includegraphics[scale=1]{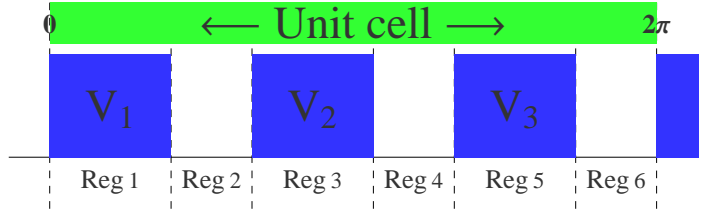}
\caption{(Color online)~Schematic of the unit cell in which the cell is divided into different regions and some regions are characterized by the scattering
potentials $V_{1}, V_{2}, V_{3}$ chosen randomly along the cell.}
\label{cell}
\end{figure}
%---------------------------------------------------------------------------------
%
%---------------------------------------------------------------------------------
Now we divide the ring into different regions in space (see Fig.~\ref{cell}) and define matrix $\Omega_n(\varphi) = (\Psi_n^1(\varphi), \Psi_n^2(\varphi))^T = C_n W_n(\varphi)$ 
for the $n^{th}$ region~\footnote{Here we have omitted the index $l$ for $\Psi_n(\varphi)$, $\Omega_n(\varphi)$ and $W_n(\varphi)$.}, where
\begin{eqnarray}
&&\Psi_n^{1,2}(\varphi) = A_n^{1,2}\psi^+_l(\varphi) + B_n^{1,2}\psi^-_l(\varphi) \ , \\
&& C_n = \left( \begin{array}{cc} A_n^1 & B_n^1 \\ A_n^2 & B_n^2 \end{array} \right)\ ;
\hspace{1cm} W_n(\varphi) = \left( \begin{array}{cc} \zeta_1^+ & \zeta_2^+ \\ \zeta_1^- & \zeta_2^- \end{array} \right)\ .
\label{p2}
\end{eqnarray}

Hence solution for any individual region can be characterised by the corresponding coefficient matrix $C_n$. For two consecutive regions, $C_n$ will be connected by a suitable 
transfer matrix such that $C_{n+1}=T_n\cdot C_n$~\cite{McKeller}. By applying this transfer matrix method sequentially in every region of the ring one can find $T_n$ 
for every region in terms of the corresponding $W_n(\varphi)$ from the continuity of $\Omega_n(\varphi)$. Now, if a ring has $N$ number of such distinct regions in space, 
then the transfer matrix for the entire ring can be defined as $C_{N+1}=T_R C_1$ where $T_R =T_NT_{N-1}\cdot\cdot\cdot T_1$. Combining the latter with the periodic boundary 
condition $\Psi_l^{1,2}(2\pi) =e^{\pm i 2\pi \Phi/\Phi_0} \Psi_l^{1,2}(0)$~\cite{Gefen1} we finally obtain our energy band equation as
\begin{eqnarray}
&&2\cos(2\pi \Phi/\Phi_0) = Tr[T_R]\ .
\label{band}
\end{eqnarray}    
where in Eq.~(\ref{band}) the role of the Bloch vector is played by the magnetic flux. Hence, the \pcd can be obtained from the band structure according to the 
well known formula~\cite{Gefen1,Gefen2}
\begin{eqnarray}
&& I_{P}=-c \frac{\partial E(\Phi)}{\partial \Phi}\ .
\label{pcf}
\end{eqnarray}

The same model can be easily modified for a one dimensional ring made of monolayer graphene by replacing the velocity of light ($c$) with the Fermi velocity 
($\rm v_F \sim 10^6 m/s$)~\cite{Peet}. In the following sections we will use $v_F$ which would be helpful to obtain a quantitative estimation of the \pcd for condensed matter systems.

%---------------------------------------------------------------------------------------------------
\section{\label{sec:rel} Single scattering potential}
%-------------------------------------------------------------------------------------------
In this section we discuss the energy spectrum and the \pcd for massive relativistic Dirac electrons on a ring in presence of single scattering potential with various height and width. 

%-----------------------------------------------------------------
\subsection{Effect of mass gap and height}
\begin{figure}[ht]
\centering
\epsfig{file=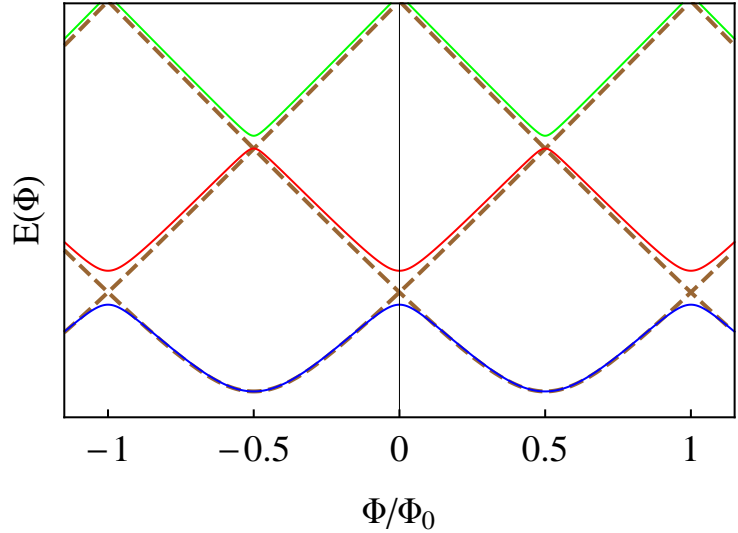,scale=1}
\caption{(Color online)~Energy spectrum for the relativistic particle with $\rm \delta=14meV$ as a function of the flux without (dashed line) as well as with (solid line) 
single scattering potential of strength $\rm V_{0}=35meV$. The radius of the Dirac ring is chosen to be $\rm R=10nm$.}
\label{energy}
\end{figure}

%-----------------------------------------------------------------
Even in absence of any scatterer the energy spectrum of a relativistic Dirac particle is not simply a quadratic function of the flux $\Phi$ \cite{Peet}. For large mass the 
dependence appears to be approximately quadratic (See Fig.~\ref{energy}) and its change due to the addition of a single scatterer is similar to the nonrelativistic 
case~\cite{Gefen1,Gefen2}. Here we have chosen positive scattering potential ($V_0>0$) for our analysis. Similar behaviour can be observed for a negative scattering potential ($V_0<0$) 
which will be discussed later.
%--------------------------------------------------------------------------------------------------
%
%--------------------------------------------------------------------------------------------------
%-----------------------------------------------------------------------------------------------------
%
%-----------------------------------------------------------------------------------------------------

\begin{figure}[h]
\centering
%\subfigure[]{\includegraphics[width=0.3\textwidth]{rip1d1.eps}}
%\subfigure[]{\includegraphics[width=0.3\textwidth]{rip1d2.eps}}
\begin{tabular}{cc}
\epsfig{file=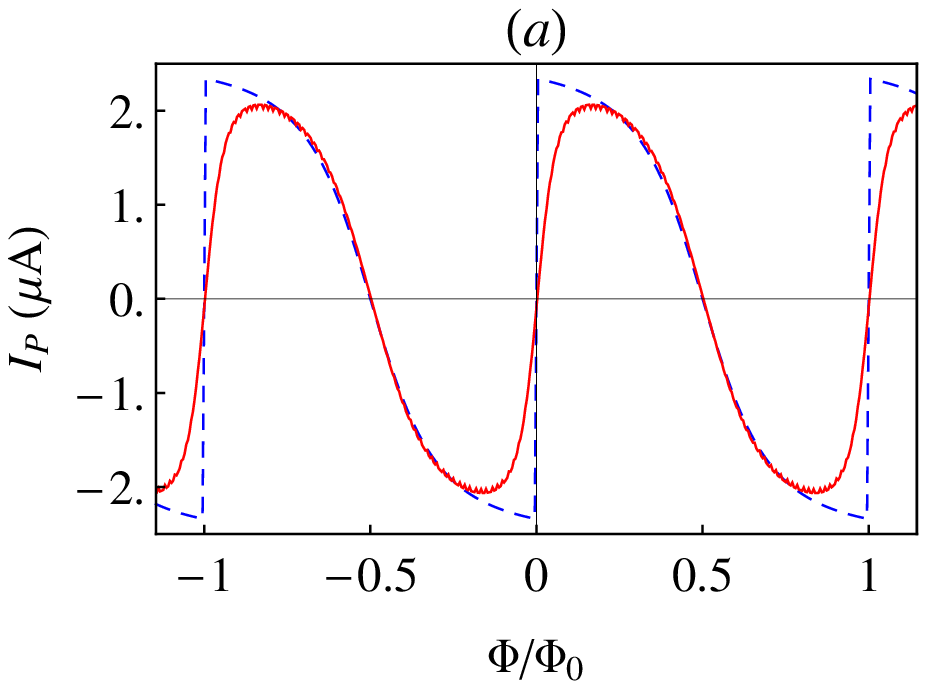,scale=0.75} &  
\epsfig{file=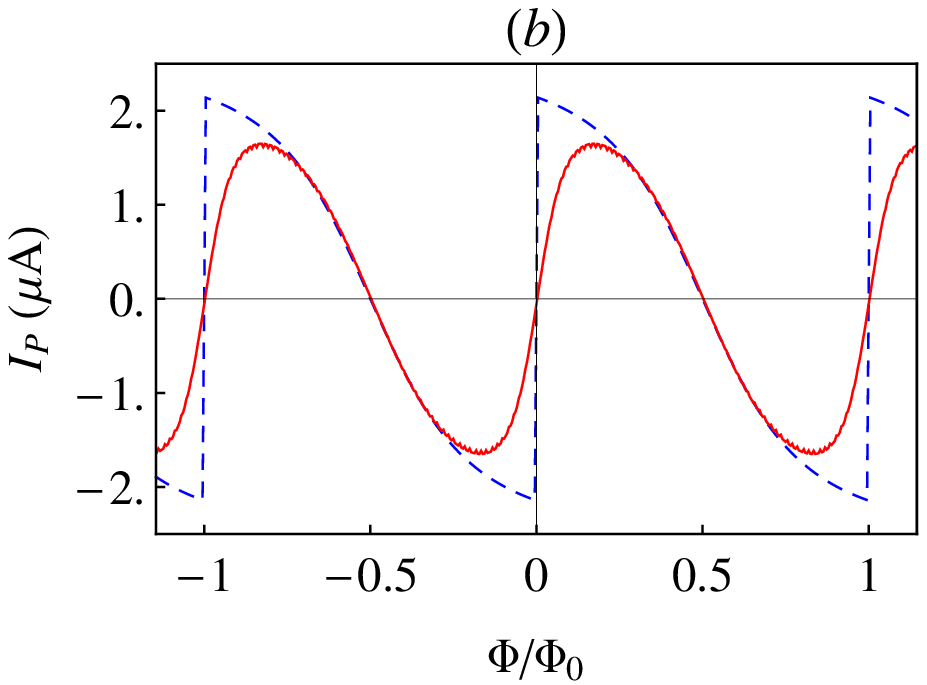,scale=0.75} 
\end{tabular}
\caption{(Color online)~Variation of \pcd as a function of the flux for massive relativistic electrons is shown for (a) $\rm \delta=14 meV$ and (b) $\rm \delta=21 meV$ without 
(Blue dashed line) and with (Red solid line) scattering potential $\rm V_0=35 meV$. The radius of the Diarc ring are chosen to be $\rm R=10nm$.}
\label{c1}
\end{figure}

In presence of a single scatterer a gap opens up in the spectrum due to which the magnitude of the maximum \pcd is reduced [see solid line of Fig.~\ref{c1}(a,b)] than the 
scattering free case [dashed line of Fig.~\ref{c1}(a,b)]. Such behavior is similar to the nonrelativistic situation where the maximum value of the \pcd always becomes smaller 
in magnitude in presence of single scattering of equal strength~\cite{Gefen1,Gefen2}. 

%----------------------------------------------------------------------------------------------
%
%---------------------------------------------------------------------------------------------
%-----------------------------------------------------------------------------------------------------
%
%-----------------------------------------------------------------------------------------------------
In Fig.~\ref{c1}(a) and Fig.~\ref{c1}(b) we show the \pcd for two different mass gaps. One can see that the amplitude of the \pcd decreases with the increase 
of the mass term $\delta$. 
From this two figures, the amplitude of maximum current is decreasing linearly with the particle mass. Keeping in mind that \pcd is also inversely proportional to the square 
of the ring radius~\cite{Papp,Su}, we see that for a ring with radius $300\rm nm$ if we substitute the particle mass by electron mass, i.e. $\rm \delta=500 eV$, 
we can roughly estimate that the current will be of the order of $0.1 \rm nA$, which is in fair agreement with the experimental finding~\cite{Jayich}. 
One can also see that the $\Phi_0$ periodicity is still there in the relativistic case with the characteristic shift of $\Phi_0/2$~\cite{Papp,Su}. 

\begin{figure}[ht]
\centering
\begin{tabular}{cc}
\epsfig{file=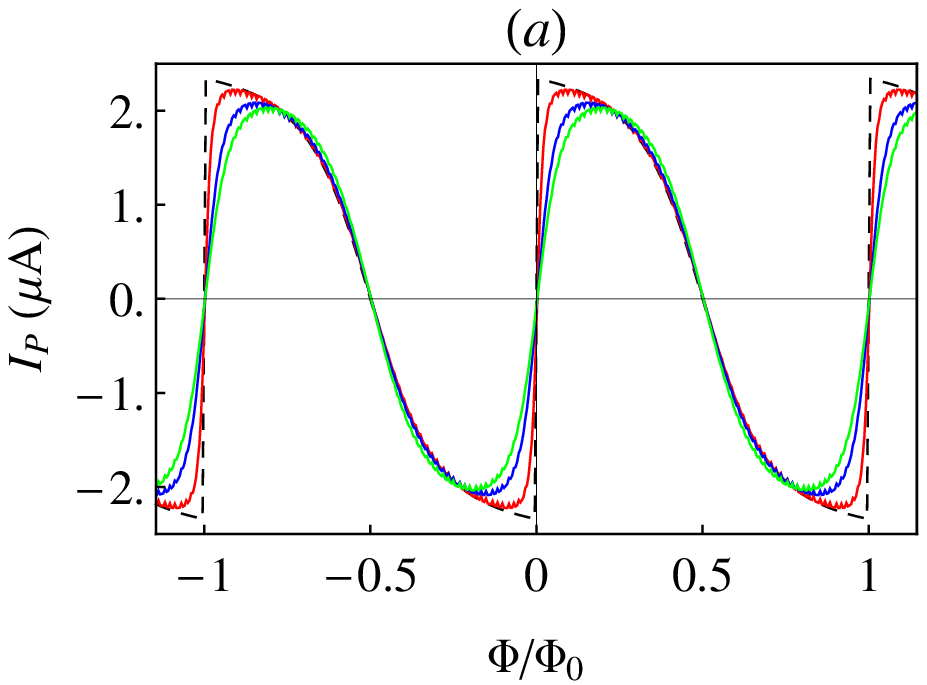,scale=0.75} &  
\epsfig{file=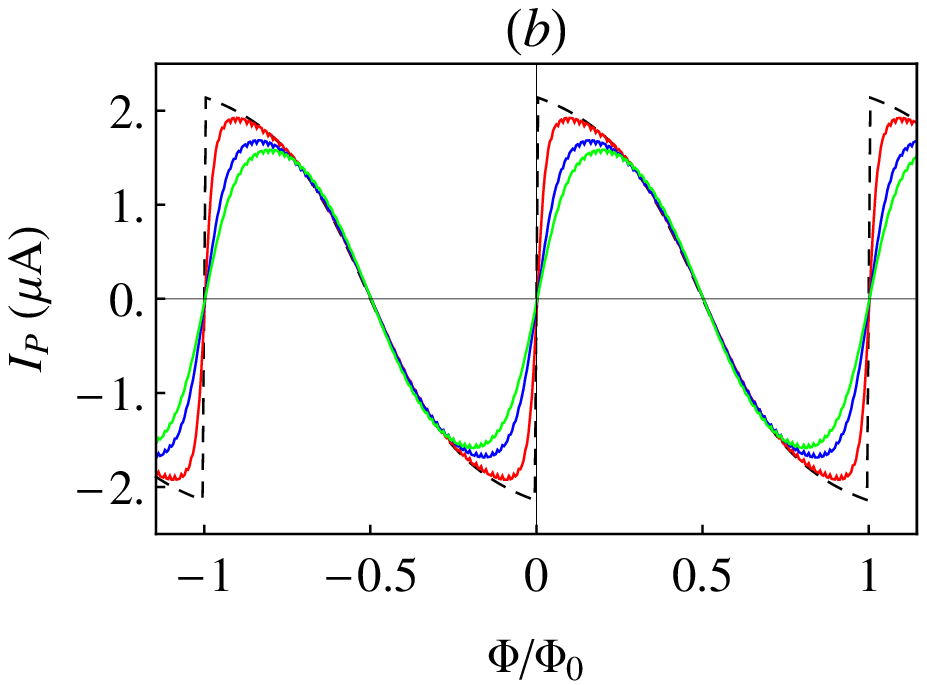,scale=0.75} 
\end{tabular}
\caption{(Color online)~Variation of \pcd as a function of the flux in presence of single scattering potential with different strengths for ($a$) $\rm \delta=14meV$ and ($b$) 
$\rm \delta=21meV$. Here different lines correspond to the attractive scattering potentials with $\rm V_{0}=10meV$ (red), $\rm V_{0}=30meV$ (blue) and $\rm V_{0}=50meV$ (green) 
respectively. The black dashed line depicts the scattering free \pc. The radius of the ring is $\rm R=10 nm$.}
\label{cv3}
\end{figure}

In Fig.~\ref{cv3}(a) and Fig.~\ref{cv3}(b) we show the behaviour of the \pcd in presence of single potential barrier with different strengths. It is clear that
the magnitude of the \pcd becomes smaller and smaller as we increase the strength of the potential which is quite expected.
%----------------------------------------------------------------------------------------
%
%-----------------------------------------------------------------------------------------
%----------------------------------------------------------------------------------------
%
%---------------------------------------------------------------------------------------
 However what distinguish it from nonrelativistic situation is that if we keep increasing the strength of the scatterer, the maximum current instead of reaching zero, 
 reaches a finite saturation value. We shall discuss about this case in the next subsection.
%-----------------------------------------------------------------
\subsection{Variation of maximum current}
\begin{figure}[h]
\centering
\epsfig{file=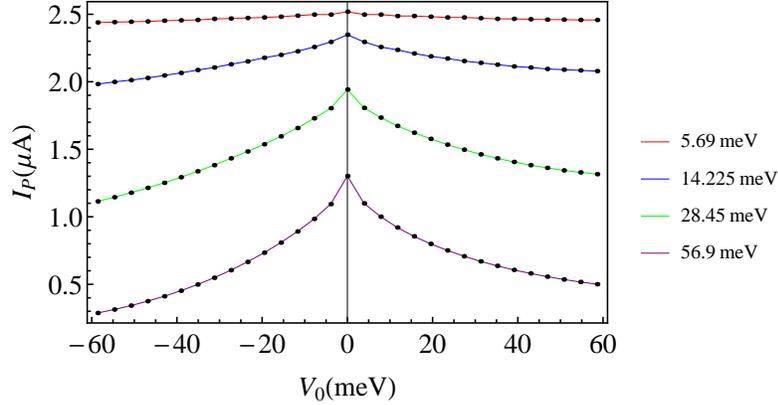,scale=0.8}
\caption{(Color online)~Variation of maximum \pcd is shown with respect to the strength of a single scattering potential for relativistic electrons with 
four different mass gaps $\delta$ (written in legend). The radius of the ring is chosen to be $\rm R=10nm$.}
\label{comparison}
\end{figure}

%-----------------------------------------------------------------
In this section we present a comparative analysis between the effect of the attractive and repulsive single scattering potential on \pcd for the relativistic case 
(see Fig.~\ref{comparison}).
%--------------------------------------------------------------------------------------------------
%
%--------------------------------------------------------------------------------------------------
%-----------------------------------------------------------------------------------------------------
%
%------------------------------------------------------------

The variation of maximum current is almost symmetric around $\rm V_{0}=0$. The asymmetry is an outcome of the fact that we have considered only positive energy particles. 
One can see that as we consider smaller values of mass term ($\delta$) the behaviour of \pcd becomes more and more symmetric. Note that in both cases the maximum magnitude 
of the \pcd decreases as we increase the strength ($\rm V_{0}$) of the potential. The relativistic \pcd remains at a finite value even in the presence of strong scattering potential. 
This is due to Klein tunneling~\cite{Setare} which allows electrons to pass through the barrier although the particle energy is less than the barrier height when the barrier 
height exceed twice the particle rest mass energy. Also note that, for $\delta \rightarrow 0$, \pcd remains almost constant, which is in complete agreement with the 
fact that massless particles can't be affected by potential barriers. 

The behaviour of maximum \pcd depicted in Fig.~\ref{comparison} can be further illustrated by the phenomena of Klein tunneling for the relativistic case.
In presence of single impurity, if the scattering potential strength $\rm V_{0} \rightarrow \infty$ and the mass term $\rm \delta \rightarrow 0$,
then the barrier becomes transparent (transmission probability through the barrier $\rm T \sim 1$) to the incident electrons due to Klein tunneling. 
In this case we obtain large and almost constant PC as shown in Fig.~\ref{comparison}.

On the other hand, if  both $\rm V_{0}$ and $\rm \delta$ are large but finite, then transmission through the barrier becomes $0<\rm T<1$ 
as shown in Ref.~\cite{Setare} and we obtain the maximum value of \pcd which is smaller in magnitude 
compared to the previous case mentioned earlier. Hence, if we increase $\rm \delta$ further (keeping $\rm V_{0}$ same as before), then $\rm T$ 
becomes smaller in magnitude than before and the corresponding \pcd is further reduced. The latter behavior is also depicted in Fig.~\ref{comparison}.
%------------------------------------------------------------
\section{\label{sec:scattering} Multiple impurities} 
%------------------------------------------------------------
Here we discuss the behaviour of \pcd for the Dirac fermions in presence of multiple scattering potentials with uniform as well as random configurations 
({quenched disorder). 
In the latter case, we assign the scattering potentials randomly on the ring for a particular configuration and consider the the average of energy/current 
over a large ($\sim1000$) number of configurations which is more realistic as far as the experimental situation is concerned. 
The corresponding mean free path ($\xi$) is of the order of typical system size ($L$) \ie~$\xi\sim L$ as in our model we consider
elastic scatterers which do not break the phase coherence of electrons.
%-----------------------------------------------------------------
\subsection{Uniform configuration}
%-----------------------------------------------------------------
Here we study a relativistic ring with scattering potentials covering 40\% of the ring for uniform spacing. In particular here we discuss the results with 3 and 10 impurities.
The uniform configuration almost overlap with the scattering free spectrum as shown in Fig.~\ref{r15}(a,c).  
Also the magnitude of the corresponding \pcd remains almost the same as the scattering free case which is shown in Fig.~\ref{r15}(b,d). 
The reason behind such behavior can be the resonance scattering inside the ring aided by the relativistic tunnelling in presence of uniform configuration of impurities.
%-----------------------------------------------------------------------------------
%
%-----------------------------------------------------------------------------------
\begin{figure}[h]
\centering
\epsfig{file=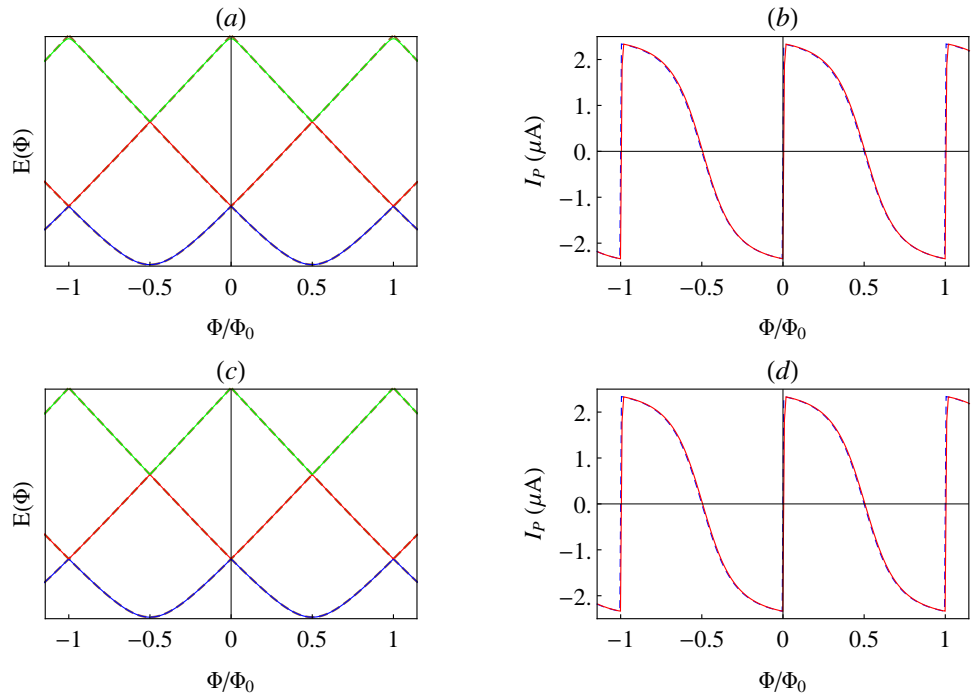,scale=0.9}
\caption{(Color online)~The energy spectrum of the quantum ring for the relativistic case is shown in (a) and (c) with uniformly spaced 3 and 10 scattering potentials respectively such that they cover 40\% size of the ring. The brown dashed line corresponds to the scattering free ring. Variation of \pcd is shown in (b) and (d). 
In all these cases, the average strength of the scattering potentials are chosen to be $\rm V_{0}=35meV$. The value of the other parameters are $\rm \delta=14 meV$ and $\rm R=10 nm$.} 
\label{r15}
\end{figure}
%------------------------------------------------------------------------------
%
%------------------------------------------------------------------------------

%-----------------------------------------------------------------
\subsection{Random configuration : Disorder average}
%-----------------------------------------------------------------
In this subsection we consider multiple impurities and study the \pcd averaged over different random configurations. For better contrast with the scattering free and 
single scatterer case, first we consider 3 scatterers alng the ring. The \pcd becomes larger in magnitude than the single scatterer case, but remains less than the scattering free scenario. 
The corresponding behaviour of disorder averaged spectrum and \pcd in presence of random scattering configurations is shown in Fig.~\ref{r3}(a) and Fig.~\ref{r3}(b) respectively.

%-------------------------------------------------------------------------------------
%
%------------------------------------------------------------------------------------
\begin{figure}[h]
\centering
\epsfig{file=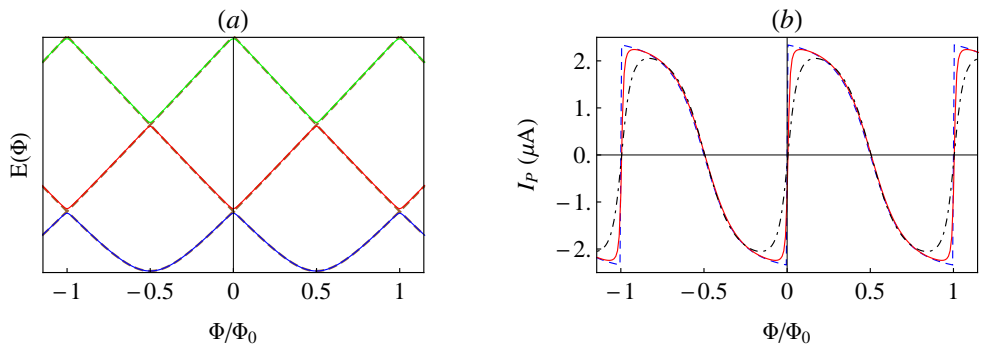,scale=0.9}
\caption{(Color online)~Disorder averaged energy $(a)$ and \pcd $(b)$ spectra for 3 scatterer which cover 40\% size of the ring. The brown dashed line in $(a)$ corresponds 
to the scattering free ring. The solid red line in $(b)$ is the disorder averaged current and the blue dashed line is the scattering free case. The single scatterer
\pcd is shown by the black dot-dashed line which is less than the 3 scatterer \pc. In all these cases, the average strength of the scattering potentials are chosen to be 
$\rm V_{0}=35meV$. The value of the other parameters are $\rm \delta=14 meV$ and $\rm R=10 nm$.} 
\label{r3}
\end{figure}
%------------------------------------------------------------------------------------
%
%-------------------------------------------------------------------------------------

Note that in our analysis we have used a finite width for our scatterer. However it can be easily generalised for a $\delta$-function potential by considering 
the zero width limit. Reduction of the width of the scatterer will produce the same effect as reducing the potential strength. If we approach to the zero width 
limit in such a way that the product of the scattering width and strength remain the same, the resulting \pcd will also exhibit the similar behavior (see Fig.~\ref{deltaV}). 
%----------------------------------------------------------------------------------
%
%----------------------------------------------------------------------------------
\begin{figure}[h]
\centering
\epsfig{file=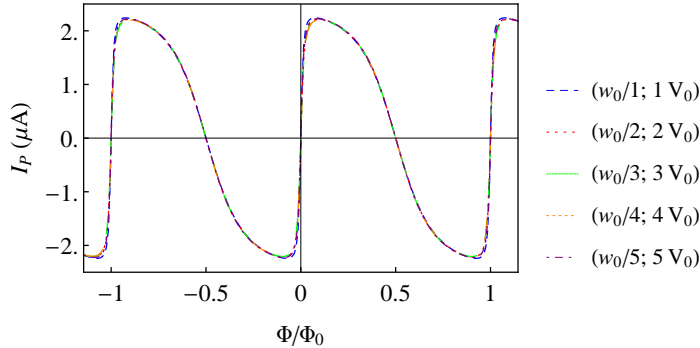,scale=0.9}
\caption{(Color online)~Disorder averaged \pcd for three scatterer configurations with different width and height of impurities such that their product remains the same. 
Individual widths and heights are written in legend. Other parameter values are  $w_0=40\%$ (total percentage of ring covered by scattering potential), 
$\rm \delta=14meV$, $\rm R=10nm$ and $\rm V_{0}=35meV$.} 
\label{deltaV}
\end{figure}
%---------------------------------------------------------------------------------------
%
%---------------------------------------------------------------------------------------

In Fig.~\ref{pcpm}(a) and Fig.~\ref{pcpm}(b) we show the disorder averaged \pcd in presence of attractive and repulsive randomly spaced scattering configurations respectively. 
The behavior of \pcd is similar for the two cases. Note that the disorder averaged \pcd is enhanced is magnitude than the single scatterer case as we increase the number of 
random configurations. This behavior is in contrast to the non-relativistic case where \pcd becomes smaller in magnitude than the single scatterer case due to the electron 
localization in presence of random impurities~\cite{Gefen1,Gefen2}. 
%The enhancement of the \pcd for the relativistic case can be a combined effect of resonance tunnelling inside the ring and relativistic tunnelling.

In presence of a scattering potential the transmissivity of a particle decreases causing a decrease in current. 
However for relativistic particle, due to Klein tunnelling, there can be a resonant transmission at an energy less than the barrier height. 
In presence of multiple scatterers, the number of resonant energy levels also increases which can cause an increase in the current under 
appropriate condition. Qualitatively, due to the presence of multiple scatterers there can be resonant tunneling through the scattering potentials 
in contrast to the single impurity.

Resonant tunneling has been studied in case of double barriers/wells in graphene in Ref.~\cite{Pereira}, considering a 
similar simplified model like us in case of graphene with a constant mass gap. Hence, some random configurations of scattering potentials
can give rise to resonant tunneling through them which is the reason for the enhancement of \pcd in the relativistic case.
On the other hand, in the non-relativistic case, random scatterers always cause the electron wave function to be localized at the scattering centres
(Anderson localization) due to which \pcd becomes smaller in magnitude as shown in Ref.~\cite{Gefen1,Gefen2}.

%----------------------------------------------------------------------------------
%
%----------------------------------------------------------------------------------
\begin{figure}
\centering
\epsfig{file=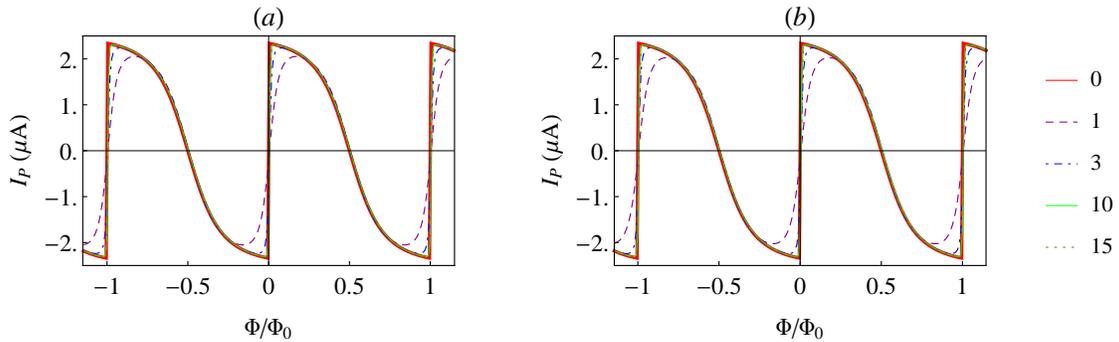,scale=0.9}
%\begin{tabular}{cc}
%\epsfig{file=Fig9a.eps,scale=0.9}&
%\epsfig{file=Fig9b.eps,scale=0.9}
%\end{tabular}
\caption{(Color online)~Disorder averaged \pcd for different number of scatterer configurations covering 40\% of total ring with (a) positive and (b) negative potential for 
$\rm \delta=14meV$, $\rm R=10nm$ and $\rm V_{0}=\pm 35meV$.} 
\label{pcpm}
\end{figure}
%---------------------------------------------------------------------------------------
%
%---------------------------------------------------------------------------------------

%-------------------------------------------------------
\section{Summary and Conclusion \label{sec:conclusion}}
%-------------------------------------------------------
To summarize, in this article we have presented the \pcd for massive relativistic electrons on a \od Dirac ring in presence of single as well as multiple attractive and 
repulsive scattering potentials. In presence of single scattering potential, the maximum \pcd for the Dirac ring decreases with the strength of the scattering with a 
rate inversely proportional to the particle mass. Even in presence of strong potential, a finite value of the \pcd remains for a massive particle. These two outcomes 
can be identified as a manifestation of Klein tunnelling. Finally in presence of multiple scattering potentials with both uniform and random configurations the \pcd is 
increased in magnitude than that of the single scatterer case after the disorder average due to resonant tunneling. This behaviour is in contrast to the non-relativistic 
case where \pcd becomes smaller in magnitude than the single scatterer case in presence of random scattering potentials.

In presence of multiple subbands (many electrons) for a ring with finite width, our results for the \pcd remain qualitatively the same
as long as the subbands are not interacting to each other. Although, if one considers band mixing, then the qualitative feature of the PC remains to be the same apart 
from the quantitative change in the current. Moreover, the period of \pcd always remains to be $2 \pi$.

The Dirac Hamiltonian we considered here is quite suitable for a monolayer graphene ring considering single valley~\cite{Peet}. As far as the practical numbers are concerned, 
according to our numerical analysis, for a typical band gap of $\rm \delta \approx 14 meV$~\cite{Lijie} and impuity strength $\rm V_{0}\approx 35 meV$, 
a \pcd of magnitude $\rm I_{P}\approx 2 \mu A$ can be obtained for a Dirac ring of radius $\rm R= 10 nm$ which can be measured experimentally.

%----------------------------------------------------------------------------------------
\section*{Acknowledgements}
%----------------------------------------------------------------------------------------
S. G. likes to acknowledge the financial support by the CSIR, India. 
The work of A. S. has been supported by the Swiss NSF, NCCR Nanoscience, and NCCR QSIT.
%\pagebreak 

%--------------------------------------------------------------------------------
\section*{References}
%---------------------------------------------------------------------------------

\end{document}

%% file: manuscript_macro.tex
\newcommand\beq{\begin{equation}}
\newcommand\eeq{\end{equation}}

\newcommand\bea{\begin{eqnarray}}
\newcommand\eea{\end{eqnarray}}

\newcommand\od{1{\textendash}D~}

\newcommand\ie{{\it{i.e.}}}
\newcommand\pc{{\textsf{PC}}}
\newcommand\pcd{\textsf{PC~}}
\newcommand\kp{{\textsf{KP}}}
\newcommand\kpd{\textsf{KP~}}
\newif\ifboo \boofalse